# Generation Expansion Equilibria with Predictive Dispatch Model

Sourabh Dalvi, David Biagioni, Muhammad Bashar Anwar, Gord Stephen, Bethany Frew

*Abstract--* This paper proposes a methodology to solve generation expansion equilibrium problems using a predictive model to represent the equilibrium in a network constrained electricity market. The investment problem for each generation company (Genco) is a bi-level problem, which is traditionally represented as a mathematical program with equilibrium constraints (MPEC), with the investment decisions in the upper level and market clearing in the lower level. The predictive model is trained using the XGBoost machine learning algorithm for estimating the system-wide revenues for each technology type across energy, operating reserves, and capacity markets in the lower level given the technology-specific installed capacity on the grid in the upper level. The profit maximization investment problem for each Genco is solved using a differential evolution global search algorithm with the predictive model. To solve for the strategic equilibrium among all Gencos, each Genco's problem is plugged into a diagonalization algorithm that is generally used in multileader, single-follower bi-level problems. The methodology presented here enables significant computational improvements while still capturing the desired market characteristics and dynamics of traditional equilibrium modeling approaches.

## Nomenclature

*Indices:*

| | |
|---|---|
| $g$ | Generation technologies |
| $j$ | Generation companies (Gencos) |
| $r$ | Load regions |
| $t$ | Operational time periods |
| $u$ | Generations units |
| $c$ | Capacity demand curve segment |

*Parameters:*

| | |
|---|---|
| $CAPEX_{r,g}^j$ | Cost of capital for generation technology $g$ in region $r$ for Genco $j$ |
| $FOM_g$ | Fixed cost of operation and maintenance for technology $g$ |
| $C_g$ | Variable cost of production for technology $g$ |
| $\theta_{r,g}$ | Predictive model function for region $r$ and technology $g$ |
| $K_{r,g}^{max}$ | Maximum investment limit for region $r$ and technology $g$ |
| $D_t$ | Demand for time period $t$ |
| $K_{r,g}^{j0}$ | Existing installed capacity for Genco $j$ in region $r$ and technology $g$ |
| $A_c^{cap}$ | Capacity demand curve price for segment $c$ |
| $M_c^{cap}$ | Capacity demand curve slope for segment $c$ |
| $D_c^{cap}$ | Maximum capacity demand for segment $c$ |
| $B_u^{cap}$ | Capacity market bid for generation unit $u$ |
| $P_u^{max}$ | Maximum power generation limit for generation unit $u$ |
| $\Delta_u$ | Derating factor for generation unit $u$ |

*Variables:*

| | |
|---|---|
| $q_{g,t}$ | Energy production from technology $g$ at time period $t$ |
| $q_t^{unmet}$ | Unmet energy demand at time period $t$ |
| $\lambda_t^e$ | Energy price dual at time period $t$ |
| $x_{r,g}^j$ | Investment strategy for Genco $j$ in region $r$ and technology $g$ |
| $k_{r,g}$ | Total installed capacity for region $r$ and technology $g$ |
| $d_c^{cap}$ | Cleared capacity demand from segment $c$ |
| $m_u^{cap}$ | Cleared capacity of generation unit $u$ |

## I. Introduction

### A. Background

Deregulation of electricity markets has challenged capacity expansion planning methods in the electricity industry. Unbundling of generation and transmission makes generation expansion planning much more competitive for self-interested, independent participants in electricity markets. Traditional planning studies often employ large-scale capacity expansion models [1], which implicitly assume a centralized planner (i.e., a vertically integrated utility) minimizing total system costs (or maximizing total social welfare) subject to constraints that ensure reliability in different market clearing scenarios. These models do not incorporate investor preferences or profit-maximizing objectives. However, in an electricity market, each generation company (Genco) can pursue its own payoff maximization by strategically planning and operating (within market power rules) its generating units. This leaves the market operator responsible for satisfying operational reliability with available supply bids.

Of the existing decentralized planning models, bi-level equilibrium models based on game-theory principles have recently gained attention. Equilibrium-based analysis is useful for gaining insights into the strategic investment behavior between producers and the evolution of generation investment. Such insights may allow the market regulator to design rules that increase market competition and economic efficiency.

A bi-level problem is defined as a mathematical optimization problem that is constrained by another optimization problem. A mathematical program with equilibrium constraints (MPEC) is an optimization problem with equilibrium constraints that are defined by a variational inequality or a complementarity constraint, which typically model a certain equilibrium condition and are often used to study bi-level problems. An equilibrium program with equilibrium constraints (EPEC) is a mathematical problem that finds an equilibrium point that satisfies the Karush-Kuhn-Tucker (KKT) optimality conditions of several MPECs simultaneously. These problems are known to be NP-hard problems even in the simplest form (i.e., when variables are continuous and constraints are linear). Even proving a solution is locally optimal can be NP-hard. In most cases, the MPEC is reformulated as a mixed integer linear program (MILP), mixed complementarity model, variational inequality problem, or nonlinear programming problem [2].

*B. Literature Review and Contributions*

Equilibrium-based models have been increasingly applied to generation expansion problems in deregulated electricity markets. Most spot markets use a unit commitment (UC) model, which is a MILP that presents serious computational tractability challenges for analysis with strategic behavior of participants either in market operations or investment. We highlight here examples of existing implementations of bi-level modeling techniques for studying electricity markets, and the proposed simplifications to tractably capture strategic behavior in investment decisions.

Authors in [3] propose a method to approximate closed-loop capacity equilibria using an open-loop formulation where Gencos simultaneously choose investment capacities and dispatch quantities to maximize individual profit based on a simplistic conjectured-price-response to their decisions. Similarly in [4], a bi-level model (MPEC) is proposed to help a producer make multistage generation investment decisions considering the investment uncertainty of rival producers using a conjecture-price-response model in the lower level. In [5], previous work is extended by using the conjecture-price-response MPEC to formulate EPEC by using KKT conditions. All the approaches in [3] – [5] rely on simplifying the market clearing model (based on conjectured-price-responses) to allow transformation of the MPECs and the EPEC into MILPs for avoiding bilinear terms and nonlinear constraints.

The authors in [6] present a three-level equilibrium model for generation and transmission expansion. The lower-level model represents the equilibrium of the energy market, the intermediate level represents the equilibrium in generation capacity expansion, and the upper-level depicts transmission expansion planning. The energy market clearing model assumes constant marginal cost linearly decreases with new installed capacity. The MPECs are formulated as a convex optimization problem and the resulting EPEC from the KKT conditions of the MPECs is reformulated into a MILP using binary expansion and Fortuny-Amat linearization.

The authors in [7] present a bi-level model whose upper-level problem determines the optimal investment and the supply offering curves to maximize its profit, and a lower level represents a pool-based electricity market with network constraints that uses demand blocks and bids. The resulting EPEC model is also transformed into a MILP using Big M representation of complementarity constraints and Strong Duality to replace bilinear terms.

In addition, there are several papers in the literature that focus on spot market equilibria. (e.g., [8] and [9]). Most work referenced above relies on some market clearing model simplification, which typically limits the products that can be represented (usually energy only). Even with linearization of the EPEC, the number of operation periods that can be represented is limited, as an increase in the number of binary variables can significantly increase the complexity of the MILP.

With the increase in deployment of nonsynchronous generators (e.g., wind and solar), authors in [10] and [11] raised concerns about power system reliability. To address such concerns, enhanced and new products are being introduced to signal and compensate for the full set of grid services needed for reliability. For instance, EirGrid, the transmission system operator in Ireland, expanded its procured grid services to 14 services, including 2 tertiary operating reserves and 3 ramping products [12], and it based remuneration for those services to sufficiently incentivize necessary investments [13]. Considering this context, the main contributions of this paper are that we:

1) Detail a market model capturing revenues from multiple products (e.g., energy, operating reserves, and forward capacity auctions) along with more operational periods compared to previous studies to better represent the price volatility from wind and solar generation;
2) Develop a methodology for using predictive machine learning (ML) to estimate electricity market outcomes with integer unit commitment variables while still representing strategic investment interactions of Gencos

## II. PROPOSED SOLUTION METHODOLOGY

This section describes the proposed approach, ahead of the formulation of the lower level energy and operating reserves market clearing model in Section II-B, which is followed by the forward capacity market model in Section II-C and is continued by the description of the predictive model used for capturing the relationship between installed capacity and revenues in Section II-D. Section II-E discusses the global search algorithm used in the proposed methodology.

*A. Approach*

We consider solving an EPEC problem comprised of one MPEC per Genco, where each MPEC consists of an profit maximization optimization problem to inform investment decisions, subject to a perfectly competitive market clearing problem for energy, operating reserves, and capacity. As highlighted before, an bi-level problem with a unit commitment-based formulation cannot be solved without convexifying the lower-level problem. We first propose a hybrid MPEC that approximates the market response to investment decisions using a predictive ML model. We then use a global optimization algorithm to solve for an approximate solution to the hybrid MPEC. We then solve EPEC via the diagonalization algorithm using the approximate solution for the hybrid-MPEC as proxy for the exact solution. Detailed steps of this approach are listed below and summarized in Fig. 1. Steps 1, 3, and 4 are detailed in Sections II-B through II-E. Steps 2 and 5 are described in Sections III-C and III-D.

1) **Specify market simulators**: To generate training data, we apply the production cost modeling tool from PowerSystems.jl [14] and PowerSimulation.jl [15] to a modified RTS-GMLC [17] test system to simulate the energy and operating reserves markets output. Another model is used for clearing of forward capacity auction.
2) **Sample training data:** We sample thousands of possible capacity build-outs for each technology and simulate the market outcomes using the tools from 1.
3) **Train ML model:** We use outputs from 2 to build a predictive ML model that can estimate the gross revenue for a technology in a particular region given the installed capacity of each technology in all regions.
4) **Construct approximate MPEC solver:** A reformulated hybrid MPEC is presented that uses the predictive ML models from 3 with a global optimization algorithm to find the best strategy to maximize profit for each Genco.
5) **Solve EPEC via diagonalization**: The approximate equilibrium solution to the resulting EPEC is solved using the diagonalization algorithm with the hybrid MPEC for each Genco.

*B. Production Cost Model*

To simulate market outcomes, we use Julia modeling packages PowerSystems.jl, which prepares and processes data useful for modeling electric systems, and PowerSimulation.jl, which constructs optimization models to study system operation models in steady-state. With this software, we build a market simulation with a unit commitment (UC) formulation for thermal devices and a basic dispatch formulation for hydro, wind, and PV technologies. Load balancing is done for each region, and line limits on inter-region ties are enforced, and all remaining intra-regional network constraints are relaxed. The model includes two reserve products—one each for upward and downward spinning reserves—but the proposed methodology does not restrict the number or type of products represented for future work. For the variable cost, we use a piece-wise linear function to approximate the quadratic cost function of thermal units. For simplicity, we assume combined-cycled gas turbines operate in a single mode where both the gas turbine and steam turbine are online. The case study presented in Section IV also uses a linearized version of the model where the integrality constraint on the unit commitment binary variables is relaxed.

*C. Forward Capacity Auction Model*

The capacity market clearing problem is based on the standard approach used in existing capacity markets [18] and is formulated as follows:

$$\max \sum_c A_c^{cap} * d_c^{cap} + \frac{1}{2} M_c^{cap} (d_c^{cap})^2 \qquad (1)$$
$$+ \sum_u B_u^{cap} m_u^{cap} P_u^{max}$$

s.t
$$\sum_u \Delta_u m_u^{cap} P_u^{max} = \sum_c d_c^{cap} \qquad (2)$$
$$0 \le d_c^{cap} \le D_c^{cap} \quad \forall c \qquad (3)$$
$$0 \le m_u^{cap} \le 1 \quad \forall u \qquad (4)$$

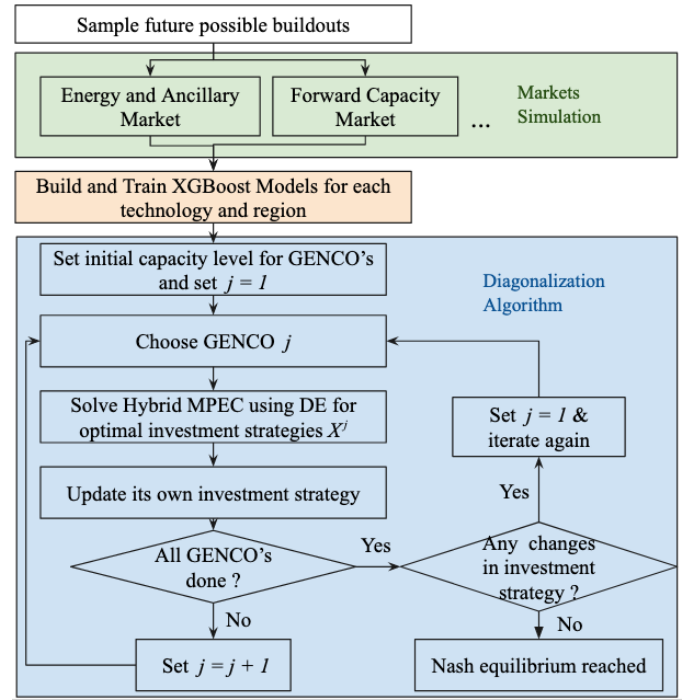

Figure 1: Flowchart of the proposed solution method.

The objective function maximizes total social welfare in capacity market, which is modeled as the difference between the welfare associated with the cleared capacity demand (based on an elastic capacity demand curve) and the cost of the capacity supply (based on market bids $B_j^{cap}$ of participating projects). The capacity balance constraint (2) ensures the total derated capacity of the cleared projects equals the total cleared capacity demand. The dual variable associated with the capacity balance constraint represents the capacity market clearing price. And (3) and (4) keep cleared demand and supply capacities within their respective limits.

*D. Predictive ML Model*

Many ML algorithms could be trained for the task of estimating the revenue based on installed capacity. In this work, we use the XGBoost algorithm, which is a decision-tree-based ensemble ML algorithm that uses a gradient-boosting framework that is highly scalable for model building and training [19]. XGBoost can work many different types of ML problem, but selecting the model parameters that can achieve a high level of accuracy is crucial. The XGBoost algorithm is highly effective and a widely used ML method [20]. The algorithm has won several data science competitions and has been credited for being the driving force for inventive industry applications.

We build an XGBoost model for each technology class and region using the results from market simulations across the sample data of thousands of different system build-outs. Revenues generated from the simulations, minus all operational costs, are set as the operational profit for each generator. For training data, we use the cumulative operational profits for each technology class and region. The input vector is the vector of installed capacities for each region and technology; the output variable is the cumulative profits for a region and technology.

### E. Global Optimization Algorithm

Evolutionary algorithms are global search algorithms inspired by biological evolution and characterized by the sequential improvement of a population of (suboptimal) solutions via selection, mutation, recombination and reproduction. We demonstrate our proposed solution technique with a differential evolution (DE) algorithm [21]. Such methods, commonly known as metaheuristics, make few or no assumptions about the problem being optimized and can search very large spaces of potential solutions, but they do not guarantee global optimality.

Due to the bi-level and very nonlinear nature of the relevant MPECs, even specialized optimization routines used to solve such problems fail for even modest problems. Our preliminary experiments, for example, show integer-based formulation would significantly slow down with 4 or more investors, 5 or more technologies, and 60 or more time periods. In contrast, once a predictive model for the MPEC objective is built, evaluating the function for a set of investment decisions is computationally cheap. So, we cast the solution of the hybrid MPEC as a global, black-box, gradient-free optimization problem that requires only objective function evaluations to work, and use a suitable algorithm, in this case the DE, to solve the hybrid MPEC problem. We note that many algorithms exist for this purpose; our choice is both convenient and effective enough to demonstrate the efficacy of the overarching approach.

## III. MPEC AND EPEC

This section details the original bi-level problem formulated as an MPEC, a linearized version used for validation purposes, the hybrid MPEC resulting from the proposed predictive ML model methodology, and the solution approach for both. In all cases, model formulations are derived from the same bi-level problem structure: the upper level maximizes profit while making strategic investment decisions and the lower level represents market equilibrium.

### A. Benchmark MPEC

To validate our proposed predictive ML model methodology, we create a linearized benchmark version of the original bi-level problem that is computationally tractable. We then compare it to a similarly linearized hybrid MPEC created with our proposed methodology. We then apply both to a stylized power system (Section IV-A).

The linear benchmark version of this upper-level investment model is shown in (5)-(6); the lower-level economic dispatch model is shown in (7)-(10); the latter minimizes operational cost subject to a demand balance constraint and production limits on $q_{g,t}$. The capacity investment variable is continuous to maintain linearity. A penalty ($VoLL$) for unmet demand $q_t^{unmet}$ is set at $1,000/MWh.

For our validation exercise (Section IV-A), we formulate the linear MPEC as a MILP by transforming the complementarity constraints using the Big M method. We substitute the bi-linear terms in the objective function with linear expressions using strong duality. For this validation, we solve the resulting EPEC using the diagonalization algorithm described in Section III-C.

Upper-level problem:
$$\max_{x_g} \sum_t (\lambda_t - C_g) q_{g,t} \quad (5)$$
$$- \sum_g \left[ FOM_g \left( K_g^{j^0} + x_g^j \right) + CAPEX_g x_g^j \right]$$

s.t
$$0 \leq x_g^j \leq K_g^{max} \quad \forall g \quad (6)$$

Lower-level problem:
$$\min_{q_{g,t}} \sum_t C_g q_{g,t} + VoLL \cdot q_t^{unmet} \quad (7)$$

s.t
$$\sum_g q_{g,t} + q_t^{unmet} = d_t \quad \forall t \quad (8)$$
$$0 \leq q_{g,t} \leq \left( K_g^{j^0} + x_g^j \right) \quad \forall g,t \quad (9)$$
$$0 \leq q_t^{unmet} \quad \forall t \quad (10)$$

### B. Hybrid MPEC and EPEC

The hybrid MPEC used in the full nonlinear UC analysis (Section IV-B) is derived from the same bi-level model (Section III-A). In this case, investment decisions are discrete, as generation units for each technology are available in fixed incremental sizes. To reduce the problem's complexity, we assume continuous investment variables as done in [3]- [5], [7], [9]. This simplifies the predictive model's input space and empirically improves performance when training data are few. From a production cost modeling perspective, aggregating the new capacity into a single unit leads to undesired commitment schedule impacts. Consequently, the continuous investment decision is split into discrete units based on a standard size for that technology class, with at most one undersized unit.

Traditional modelling approaches rely on KKT conditions or primal-dual formulation with strong duality of the lower-level market clearing problem to link the two optimization problems into a MPEC [2]. These methods work well but require the lower-level problem to be linear, which limits us to a simplified, energy-only market that ignores unit commitment constraints. We propose both the investor problem and the market clearing problems be kept in their natural form but use the XGBoost model to build link between the two problems, where the price times quantity term ($\lambda_{r,t} * q_{g,r,t}$) that traditionally is derived from the lower level is replaced by a set of predictive ML model ($\theta_{r,g}$) using the XGBoost algorithm as shown in (11)-(13).

$$\max \sum_{r,g} \left[ \theta_{r,g}(K_{r,g},\ldots,K_{R,G}) \frac{\left(K_{r,g}^{j0} + X_{r,g}^j\right)}{K_{r,g}} \right. \quad (11)$$
$$- CAPEX_{r,g}^j X_{r,g}^j$$
$$\left. - FOM_g \left(K_{r,g}^{j0} + X_{r,g}^j\right) \right]$$

s.t
$$K_{r,g} = \sum_{j \in \{J|\, i \notin J\}} K_{r,g}^j + K_{r,g}^{i0} + X_{r,g}^i \quad \forall r,g \quad (12)$$
$$0 \leq X_{r,g}^j \leq K_{r,g}^{max} \quad \forall r,g \quad (13)$$

XGBoost functions are known to be excellent general-purpose regressors, which was confirmed in our preliminary testing, where they had the best out-of-sample prediction accuracy of all methods. However, their tree-based structure means they are not amenable to gradient-based optimization. Thus, we use them as objective function surrogates and employ black-box optimization using DE to search for their global maxima. The EPEC can now be expressed as the problem of finding a Nash equilibrium among the Gencos, whose strategy variable is $X^j$, which is a vector of decision variables $X_{r,g}^j$:

$$\text{Find: } \{X^1, \ldots, X^J\} \quad (14)$$
$$\text{st. Hybrid MPEC (11)} - (13) \quad \forall j = 1, \ldots, J$$

### C. Algorithm for the Diagonalization

Diagonalization is the most common strategy for finding a solution to an EPEC. If under diagonalization, no Genco desires to deviate from its strategy, the strategies of all producers satisfy the definition of Nash equilibrium as follows:

1. Diagonalization starts by setting the initial installed capacity values for all Gencos $K_{r,g}^{j0}$, for all regions and technologies, a convergence criterion ε, and a maximum number of iterations N. The second superscript refers to the iteration of the procedure.
2. Initialize iteration counter, $n \leftarrow 1$
3. For $j = 1, \ldots, J$:
   a. Genco $j$'s MPEC is solved for its optimal strategy $X^{jn}$, given the other Gencos decisions $X^{j'} \; \forall j' \neq j$.
   b. Set $X^j = X^{jn}$
4. If $|X^{jn} - X^j| < \epsilon$ for all j, then Nash equilibrium is found, quit and report the last solution $\{X^1, \ldots, X^J\}$. Else
5. If $n = N$, quit and report the algorithm has failed to converge. Else $n \leftarrow n + 1$ and return to Step 3.

### D. Training Data and Computational Issues

Generating the training data, most computationally intensive part of our proposed methodology, requires solving a UC problem. To generate training data for the case study in Section-IV, we simulate a year's dispatch schedule for each sampled build-out, which can take 2–4 hours for each sample. A representative set of days can be used in the production cost model to reduce the solve time for a larger system. With help from warm-starting the unit commitment variables, along with high-performance computing to parallelize the market simulation, we simulate 10,000 samples in 72 hours. Crucial to the proposed method are the predictive model and its accuracy at predicting revenues. We use prediction accuracy on a held-out data subset of 25% of the training data to determine when model accuracy is saturated (i.e., no more improvement is possible).

Considering that a global search algorithm like DE can terminate at a local optimum, we run DE multiple times with randomized initial values and keep the best solution. This seeding strategy is parallelizable, but the solution time for a single instance increases with the numbers of technologies and regions. In addition, the global optimization problem generally gets more difficult to solve as dimensionality increases. Consequently, even multi-started solutions are likely to be only locally optimal. The global optima will still remain elusive with EPECs, but our proposed method provides an approximate solution to a problem that is NP-hard in its original form.

### IV. CASE STUDY

In this section, 1) we apply the proposed methodology to a numerical example and compare the method with the benchmark solution using a stylized system, and 2) we demonstrate the scalability and flexibility of the presented method on an instance of EPEC, where the benchmark methods fail to provide a solution. We also present numerical results, first from a stylized power system and then from a modified version of the IEEE Reliability Test System (RTS) from [17].

### A. Case Study 1: Hybrid EPEC validation

A stylized power system is used in the first case. It consists of three Gencos, and each can invest in three different technologies: steam turbines (ST), single cycle gas turbines (CT), and combined cycle gas turbines (CCGT). The production and investment costs are provided in TABLE I. The lower-level market clearing is represented by a linear economic dispatch model, as this allows us to compare the proposed method with the benchmark EPEC. The set of 12 representative hours are shown in TABLE II. We generate 5,000 uniformly distributed samples of different buildouts of the stylized system and simulate it through the economic dispatch model from the lower level of the benchmark MPEC.

For the next step, we proceed to build the XGBoost model for each technology, by training it to predict the gross revenue for each technology given the installed capacity. In TABLE III, we see the model fit is very good, as it is estimating a linear capacity-revenue relationship for the three technology types. In contrast, Fig. 2 compares the total market payout to the installed capacity and reveals where the revenues start declining after a period of increase; this inflection point is at 1,421 MW, which is the minimum load level. Once the installed capacity exceeds 1,421 MW, fewer periods have shortages of supply and thus lower energy prices and payouts. Thus, the benchmark EPEC's equilibrium results in 1,421 MW of installed capacity, where Gencos exercise market power in the investment level to withhold installed capacity leading to higher price periods.

The equilibrium presented TABLE IV benchmark EPEC is not the most economical from Genco's preceptive, as swapping the CT capacity with CCGT would increase the investment cost slightly but significantly reduce the production cost. This is because multiple equilibria exist for this problem, and the found one depends on initial conditions and the order in which the Gencos decide. For both EPECs, the three Gencos start with zero installed capacity, and the diagonalization proceeds in numerical order.

In the algorithm's first iteration, the best investment strategy for Genco-1 and Genco-2 is to build all technologies to the max. But for Genco-3, the best strategy is to invest in ST, which does push total capacity over 1,421 MW, so in the next iteration, Genco-1 is forced to concede its CT capacity, which is replaced by Genco-3's ST, as it is cheaper to operate for the one period when demand is met. This continues until a Nash equilibrium is reached, as shown in Fig. 3. Notably, Genco-2's decision stays the same throughout the simulation, because at the first iteration little capacity is installed, and in the subsequent iteration, Genco-1 reduces its installed capacity.

In comparison, the proposed methodology reaches an equilibrium with 1,423.5 MW of installed capacity, which is within 1% of the benchmark solution. However, as shown in **Error! Reference source not found.**V, the resulting buildouts are different because the problem has multiple equilibria. To further validate our results, we plug the solution from the Hybrid EPEC into the benchmark EPEC and rerun the diagonalization to verify the solution, which terminates at a very similar equilibrium but with slightly less capacity at 1,421 MW as expected.

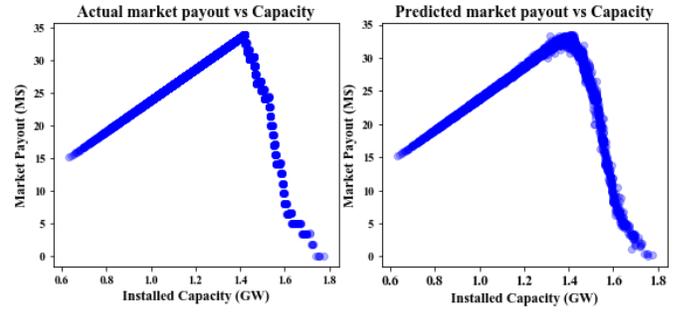

Fig. 2. Comparing market payout with the economic dispatch model (left) vs. XGBoost model (right)

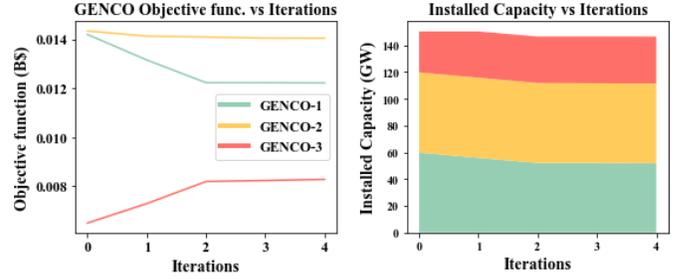

Fig. 3. Comparing objective function of each Genco's hybrid-MPEC and installed capacity vs. the iterations of diagonalization for the proposed method

TABLE I
PRODUCTION AND INVESTMENT COSTS

| Technology | Production Cost [$/MWh] | Investment Cost [$/MW] | Investment Limit [MW] |
| --- | --- | --- | --- |
| ST | 2.0 | 15.0 | 300 |
| CT | 3.0 | 9.9 | 200 |
| CCGT | 4.0 | 10.0 | 100 |

TABLE II
LOAD PROFILE

| Time Period | Load [MW] | Time Period | Load [MW] |
| --- | --- | --- | --- |
| 1 | 1,493 | 7 | 1,507 |
| 2 | 1,471 | 8 | 1,529 |
| 3 | 1,440 | 9 | 1,549 |
| 4 | 1,421 | 10 | 1,581 |
| 5 | 1,428 | 11 | 1,593 |
| 6 | 1,462 | 12 | 1,597 |

TABLE III
RELATIVE ERROR ON TRAIN AND TEST SETS FOR XGBOOST MODELS

| Technology | Train | Test |
| --- | --- | --- |
| CC | 1.06% | 2.36% |
| CT | 0.99% | 2.82% |
| ST | 1.28% | 2.9% |

TABLE IV
EQUILIBRIUM IN CASE 1

| GENCO | Benchmark EPEC Capacity [MW] | | | Hybrid EPEC Capacity [MW] | | |
| --- | --- | --- | --- | --- | --- | --- |
| | ST | CCGT | CT | ST | CCGT | CT |
| 1 | 300 | 200 | 18 | 293.4 | 196.9 | 0.4 |
| 2 | 300 | 200 | 100 | 299.9 | 189.55 | 9.1 |
| 3 | 300 | 3 | 0 | 294.6 | 133.0 | 6.30 |

*B. Case Study 2: RTS-GMLC System*

For a second case study, we use a reduced form of the RTS GMLC system [16] with a set of generating units removed to produce training data. In this case, we have seven different technologies: ST, CT, CCGT, hydro-powered turbines (HY), wind turbine (WT), battery-powered energy storage (BA) and photovoltaic (PVe). For newly built WT and PVe units, we assume the wind and solar resource are similar to that of existing generation, so energy production time series provided in the RTS dataset are scaled according to the nameplate capacity of newer units. We consider 8,760 time periods for simulating the energy and operating reserves market outcomes with a UC formulation for the target year and the forward capacity model from Section II-C. Solving the benchmark EPEC for this exact problem is impossible as the lower level is nonlinear. Even after excluding unit commitment constraints, reserve and capacity products from the formulation, solving the problem to optimality would still require unreasonable amounts of time and computational resources.

To provide an approximate solution, we start by sampling continuous investment decisions for each technology and region. We then discretize these decisions into a set of generating units for the UC formulation. We draw 10,000 samples of investment decisions uniformly at random on closed intervals, which are selected to simulate over- and under-capacity scenarios. After adding the new units to the modified RTS-GMLC system, we simulate the target year using the day-ahead energy, upward- and downward-spinning reserve load profiles. Along with the production cost model, the forward capacity market model from Section II-B simulates the results from the capacity auction. The investment parameters from TABLE V are used in each Genco hybrid MPEC, and the penalty for unmet energy and reserve demand are set at $10,000/MWh and $1,000/MWh respectively.

TABLE V
INVESTMENT PARAMETERS

| Technology | Annualized CAPEX [$/MW] | Discount Rate | Lifetime [yrs.] | Derating Factor |
|---|---|---|---|---|
| CC | 78,962.74 | 0.06229 | 20 | 0.9 |
| CT | 72,122.76 | 0.06229 | 20 | 0.9 |
| PVe | 83,448.85 | 0.04903 | 20 | 0.1 |
| WT | 104,943.38 | 0.05383 | 20 | 0.19 |
| ST | 313,093.74 | 0.06229 | 20 | 0.90 |
| Hydro | 483,931.29 | 0.05783 | 20 | 0.90 |
| BA | 115,123.28 | 0.05783 | 15 | 0.65 |

TABLE VI
RELATIVE ERROR ON TRAIN AND TEST SETS FOR XGBOOST MODELS

| Technology | Region 1 | | Region 2 | | Region 2 | |
|---|---|---|---|---|---|---|
| | Train | Test | Train | Test | Train | Test |
| CC | 2.9% | 4.7% | 2.8% | 4.7% | 2.8% | 4.7% |
| CT | 2.2% | 3.9% | 2.1% | 3.8% | 2.2% | 3.9% |
| PVe | 3.5% | 5.9% | 3.4% | 5.8% | 3.3% | 5.5% |
| WT | 9.2% | 13.1% | 9.3% | 13.2% | 8.3% | 11.6% |
| ST | 2.9% | 4.8% | 2.9% | 4.8% | 2.9% | 4.7% |
| Hydro | 3.0% | 4.9% | 3.0% | 4.9% | 2.9% | 4.8% |
| BA | 3.2% | 5.4% | 3.3% | 5.5% | 3.3% | 5.4% |

For our purposes here, all computation is done on the NREL HPC Eagle, and each node is equipped with Dual Intel Xeon Gold Skylake 6154 (3.0 GHz, 18-core) processors with 96 GB of memory. Each run of the PCM takes 2–4 hours of wall-time using a FICO Xpress optimizer. The execution time depends on the number of binary variables in the model, which is directly proportional to the number of thermal units.

The errors listed in TABLE VI show an out-of-box XGBoost model can approximate most technologies revenue very well. From Fig. 4 and Fig. 5, ST units that mostly satisfy base load in the system are clearly the easiest to predict; and the volatile nature of wind availability along with the fact that wind is subject to curtailment at higher penetration makes the predictive task much more complex for WT. This highlights that the approximate equilibrium can only be as good as the prediction models. We observe in Fig. 6, that even though the overall capacity seems to have converged, investment strategies keep changing by small amounts at each iteration of the diagonalization algorithm. This behavior is similar to that found in the first case study, and it is caused by two factors: existence of multiple equilibria and minor errors in predicted revenues.

The diagonalization algorithm is initialized with zero installed capacity, and similar to the first case study, Genco-1 and Genco-2's strategy to install as much capacity as possible is suboptimal in later iterations. Thus, the huge drop in these Gencos' objective functions at iteration 2. After iteration 10, the equilibrium solution has 9,617 MW of installed capacity with ST contributing 30%, CC and WT around 20% each, and CT, PVe and HY 8%–10%. Batteries only make up about 2% of the installed capacity due to the simplified representation in the production cost model failing at capturing the accurate value of BA technology.

The expected payoff for each Genco (Fig. 6) is caused by high prices in the energy and reserve products set by the unmet demand penalty. For perspective, using the derating factors from TABLE V, the equilibrium found has less than 6,500-MW derated capacity for a system with peak energy demand of 8,100 MW. Such behavior is expected from the Genco in this experiment, but in reality, this unmet demand would be fulfilled by imports from other regions or a new entrant in the market.

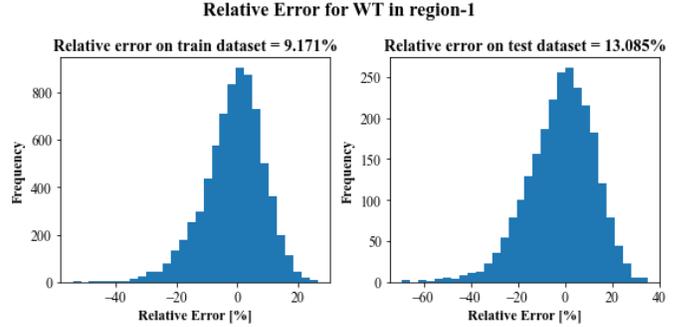

Fig. 4. XGBoost model relative error for WT in region 1 on dataset for training (left) and testing (right)

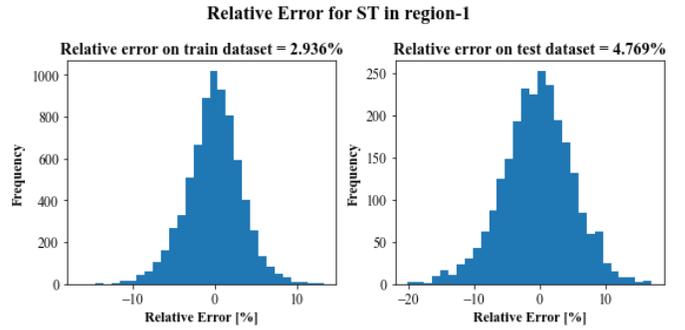

Fig. 5. XGBoost model relative error for region 3 on dataset for training (left) and testing (right)

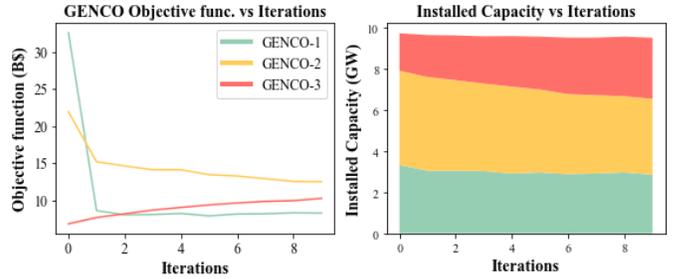

Fig. 6. Comparing objective function of each GENCO's MPEC and installed capacity versus the iterations of diagonalization.

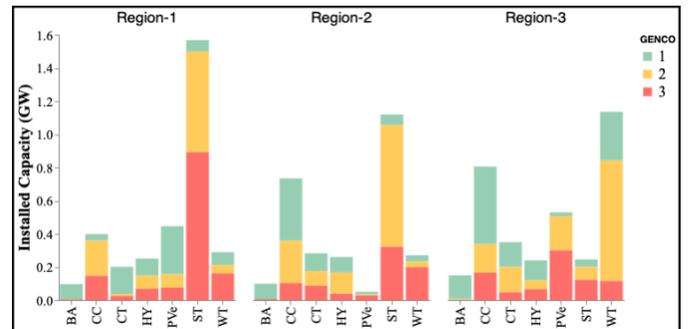

Fig. 7. Capacity build-out at equilibrium for different regions of the RTS test case.

## V. CONCLUSION

We introduce a framework for building a hybrid-MPEC model to solve generation expansion equilibria problems in deregulated electricity markets. The hybrid-MPEC optimization problem uses the XGBoost ML algorithm to anticipate the equilibrium outcomes of the lower-level problem and a DE algorithm to find optimal investment decisions. Our approximation scheme yields a strategic equilibrium with detailed representation of the energy, operating reserve, and capacity markets. We apply this approach for only one target year, but future research could extend our framework to provide an investment strategy that spans different target horizons by leveraging transfer learning to retrain the XGBoost model for the different target years.

We also conduct two case studies in which we first validate this method on a stylized numerical example against the traditional EPEC using a linearized representation of the original bi-level model. We then show the scalability and flexibility of the method on a second case study by solving for an equilibrium using very high-resolution, non-linear lower-level problem with unit commitment and multiple products.

TABLE VII
EQUILIBRIUM IN CASE STUDY 2

| Genco | Region | WT | ST | CT | CC | HY | PVe | BA |
|---|---|---|---|---|---|---|---|---|
| 1 | 1 | 112. | 183. | 119. | 29.4 | 77.3 | 198. | 94.3 |
|   | 2 | 188. | 26.5 | 75.5 | 112. | 93.8 | 9.2 | 93.3 |
|   | 3 | 527. | 20.1 | 237. | 416. | 95.3 | 47.0 | 1.48 |
| 2 | 1 | 96.9 | 865. | 44.9 | 178. | 136. | 135. | 3.1 |
|   | 2 | 191. | 822. | 189. | 492. | 124. | 0.40 | 0.57 |
|   | 3 | 505. | 133. | 68.8 | 243. | 70.8 | 216. | 0.79 |
| 3 | 1 | 84.6 | 528. | 35.5 | 193. | 42.5 | 120. | 1.4 |
|   | 2 | 64.9 | 269. | 37.6 | 131. | 35.7 | 28.1 | 5.10 |
|   | 3 | 111. | 93.9 | 38.0 | 171. | 92.6 | 252. | 0.22 |

VI. ACKNOWLEDGMENT

This work was authored by the National Renewable Energy Laboratory, operated by Alliance for Sustainable Energy, LLC, for the U.S. Department of Energy (DOE) under Contract No. DE-AC36-08GO28308. This work was supported by the Laboratory Directed Research and Development (LDRD) Program at NREL. This research was performed using computational resources sponsored by the Department of Energy's Office of Energy Efficiency and Renewable Energy and located at the National Renewable Energy Laboratory. The authors especially thank Michael Craig and Brayam Valqui Ordonez for modeling and thoughtful contributions during earlier phases of this project that ultimately evolved into this work. The views expressed in the article do not necessarily represent the views of the DOE or the U.S. Government. The U.S. Government and the publisher, by accepting the article for publication, acknowledges that the U.S. Government retains a nonexclusive, irrevocable, worldwide license to publish or reproduce the published form of this work, or allow others to do so, for U.S. Government purposes.